\documentclass{emulateapj}

\shortauthors{MINEZAKI ET AL.}
\shorttitle{Mid-IR Imaging of Quadruple Lenses}

\slugcomment{To appear in the Astrophysical Journal}

\begin{document}

\title{Subaru Mid-infrared Imaging of the Quadruple Lenses. II.
Unveiling Lens Structure of MG0414$+$0534 and Q2237$+$030
\altaffilmark{1}}

\author{Takeo~Minezaki\altaffilmark{2},
        Masashi~Chiba\altaffilmark{3},
        Nobunari~Kashikawa\altaffilmark{4},
        Kaiki~Taro~Inoue\altaffilmark{5},
    and Hirokazu~Kataza\altaffilmark{6}}

\altaffiltext{1}{Based on data collected at Subaru Telescope,
which is operated by the National Astronomical Observatory of Japan.}
\altaffiltext{2}{Institute of Astronomy, School of Science, University of
Tokyo, Mitaka, Tokyo 181-0015, Japan; minezaki@ioa.s.u-tokyo.ac.jp}
\altaffiltext{3}{Astronomical Institute, Tohoku University,
Aoba-ku, Sendai 980-8578, Japan; chiba@astr.tohoku.ac.jp}
\altaffiltext{4}{National Astronomical Observatory, Mitaka, Tokyo 181-8588,
Japan; n.kashikawa@nao.ac.jp}
\altaffiltext{5}{School of Science and Engineering, Kinki University,
Higashi Osaka 577-8502, Japan; kinoue@phys.kindai.ac.jp}
\altaffiltext{6}{Institute of Space and Astronautical Science,
Japan Aerospace Exploration Agency, Sagamihara, Kanagawa
229-8510, Japan; kataza@ir.isas.jaxa.jp}

\begin{abstract}
We present mid-infrared imaging at 11.7 $\mu$m for the quadruple lens systems,
MG0414$+$0534 and Q2237$+$030, using the cooled mid-infrared camera and
spectrometer (COMICS) attached on the Subaru telescope.
MG0414$+$0534 is characterized by a bright pair of lensed images \rm{(A1, A2)}
and their optical flux ratio A2$/$A1 deviates significantly from the
prediction of a smooth lens model. 
Q2237$+$030 is ``the Einstein cross'' being comprised of four lensed images,
which are significantly affected by microlensing in a foreground lensing galaxy.
Our mid-infrared observations of these lensed images have revealed that
the mid-infrared flux ratio for A2$/$A1 of MG0414$+$0534 is nearly unity
($0.90 \pm 0.04$). We find that this flux ratio is systematically small,
at 4 to 5~$\sigma$ level, compared with the prediction of a best smooth lens model
($1.09$) represented by a singular isothermal ellipsoid and external shear.
The smooth lens model which also considers the additional lensing effect of
the possible faint satellite, object X, provides yet a large flux ratio of
A2$/$A1$=1.06$, thereby suggesting the presence of more substructures
to explain our observational result.
In contrast, for Q2237$+$030, our high signal-to-noise observation indicates
that the mid-infrared flux ratios between all the four images of
Q2237$+$030 are virtually consistent with the prediction of
a smooth lens model.
Based on the size estimate of the dust torus surrounding the nuclei of these
QSOs, we set limits on the mass of
a substructure in these lens systems, which can cause anomalies
in the flux ratios. 
For MG0414$+$0534, since the required mass of a substructure inside its Einstein
radius is $\ga 360 M_\odot$, millilensing by a CDM substructure is most likely.
If it is modeled as a singular isothermal sphere, the mass inside radius
of 100~pc is given as $\ga 1.0 \times 10^5 M_\odot$.
For Q2237$+$030, there is no significant evidence of millilensing, so
the reported anomalous flux ratios in shorter wavelengths are entirely
caused by microlensing by stars.
\end{abstract}


\section{INTRODUCTION}

Although the cold dark matter (CDM) scenario is now a leading theoretical
paradigm for understanding the formation of large-scale structures
in the Universe, it confronts yet various discrepancies with observations
on smaller, galactic and sub-galactic scales. One of the most serious issues
is so-called ``the missing satellites problem'', namely,
while the total number of known Milky Way satellites is about twenty,
CDM models predict the existence of more than several hundred substructures of
dark matter or ``CDM subhalos'' with masses of $M \sim 10^{7-9}$ M$_\odot$,
in a galaxy-sized halo with $M \sim10^{12}$ M$_\odot$ 
(Klypin et al. 1999; Moore et al. 1999; see also Diemand et al. 2007
 for more recent, high resolution studies).
The abundance of CDM halos on small scales
is closely relevant to the small-scale power spectrum of initial density
fluctuations, thereby affecting the formation of the first generation
objects and the resultant reionization history of the Universe.
Also, their abundance in a galaxy-sized halo reflects
the merging history and dynamical evolution of dark-matter halos in the
expanding Universe, thereby being associated with the formation of
luminous galaxies within the framework of hierarchical galaxy formation
scenarios.

To clarify this outstanding issue, gravitational lensing offers us
an invaluable insight into the spatial structure of a galaxy-sized
dark halo, which works as a lens for a remote source, such as a QSO.
In particular, flux ratios between multiple images of a lensed source
are a sensitive probe for the mass distribution of a lens.
There exist a class of lensed QSOs having anomalous flux ratios, namely,
those hardly reproduced by any lens models with a smooth density distribution
and such flux anomalies can be caused by any substructures that reside
in a lensing galaxy through either of {\it millilensing} by CDM subhalos
or {\it microlensing} by stellar objects
(e.g., Mao \& Schneider 1998; Metcalf \& Madau 2001; Chiba 2002).

For the purpose of enlightening the nature of lens substructures,
mid-infrared observations of lensed images and their flux ratios provide us
with advantageous perspectives compared with the studies in other wavelengths.
Firstly, mid-infrared flux is free from extinction effects. The extinction
by intervening dust working differentially on each lensed image does not
affect flux ratios in mid-infrared band, contrary to optical flux ratios.
Secondly, mid-infrared flux is free from microlensing effects.
Observed mid-infrared flux originates basically from the dust torus of a QSO
producing near-infrared emission at the rest frame. The torus is
so spatially extended, with a radius of $\sim 1$~pc, that its flux is
unaffected by stars in the foreground lensing galaxy (microlensing) because
their Einstein radii are only $\sim 0.01$~pc. On the other hand, CDM subhalos
having much larger Einstein radii (millilensing) can affect the mid-infrared
flux from a dust torus.
Finally, the inner radius of a dust torus surrounding a central engine
can quantitatively be estimated from the QSO luminosity, based on
the technique of dust reverberation observations \citep{minezaki04}.
Then making use of an available source size for this lensing event,
it is possible to place a useful limit on the mass of CDM subhalos.
In this regard, mid-infrared observations are more advantageous than
radio ones, because of the availability of a source size for
both radio-loud and radio-quiet QSOs.

This is the second paper reporting our series of mid-infrared observations
of quadruple lenses with anomalous optical flux ratios,
using the Subaru telescope. Our first paper dedicated to PG1115$+$080 and
B1422$+$231 was published in Chiba et al. (2005, hereafter referred to
as Paper I). Here we report the results of two quadruple lenses
MG0414$+$0534 and Q2237$+$030, which were discovered by
\citet{hewitt92} and \citet{huchra85}, respectively.

MG0414$+$0534 is located at a source redshift of $z_{\rm S} = 2.639$ and
lensed by a foreground elliptical galaxy at a redshift of $z_{\rm L} = 0.9584$
\citep{lawrence95,tonry99}.
This lens system consists of four lensed images referred to as A1, A2, B, and C,
with the separation between a close pair of bright images A1 and A2
being $\sim 0.\arcsec 4$. This image configuration of A1 and A2 emerges
if the QSO is close to and inside a fold caustic provided by the foreground
lens and the flux ratio between both images is expected to be unity
for an ideally smooth lens. In MG0414$+$0534, this flux ratio has been
observed to be a function of wavelength, where A2 is found to be fainter
than A1, and this has been interpreted as a result of differential
extinction in the lens and/or of microlensing
of one or more of the images (e.g., Lawrence et al. 1995; Malhotra, Rhoads, \&
Turner 1997; Falco et al. 1997; McLeod et al. 1998; Tonry \& Kochanek 1999;
Angonin-Willaime et al. 1999; Ros et al. 2000).

Q2237$+$030 is located at $z_{\rm S} = 1.695$ and lensed by the central bulge
component of a foreground spiral galaxy at $z_{\rm L} = 0.0394$
\citep{huchra85}.
This is the famous Einstein Cross consisting of four images
A, B, C, and D. It has been known that these lensed images as measured
in optical wavelengths are strongly affected by microlensing,
so that the optical flux ratios between lensed images differ
from the predictions of a smooth lens \citep{irwin89}.
Agol, Jones, \& Blaes (2000) observed this system
in mid-infrared using the Keck telescope and reported that the mid-infrared
flux ratios may be explained by a smooth lens model within the limitation
of their data having intermediate S/N ratios
(see also Agol et al. 2001).
\citet{metcalf04} performed an integral-field spectroscopy of this system
and showed that the emission-line flux ratios such as those of 
$[$O$_{\rm III}]$ may suggest some signature of lens substructures.
Thus, further independent studies based on mid-infrared observations of both
systems are required to deduce the nature of their anomalous flux ratios.

The paper is organized as follows. In \S 2, we show the observations
of our current targets. In \S 3, the procedure of data reduction is presented to
obtain the mid-infrared fluxes and their ratios among the lensed images.
In \S 4, the implications for the current observational results are discussed
and the conclusions are drawn in \S 5. In what follows, we adopt the set of
cosmological parameters of $\Omega=0.3$, $\Lambda=0.7$, and $h=0.7$
($h \equiv H_0/100$ km~s$^{-1}$~Mpc$^{-1}$) for all relevant estimations,
as adopted in Paper I.

\begin{figure*}
\figurenum{1}
\plottwo{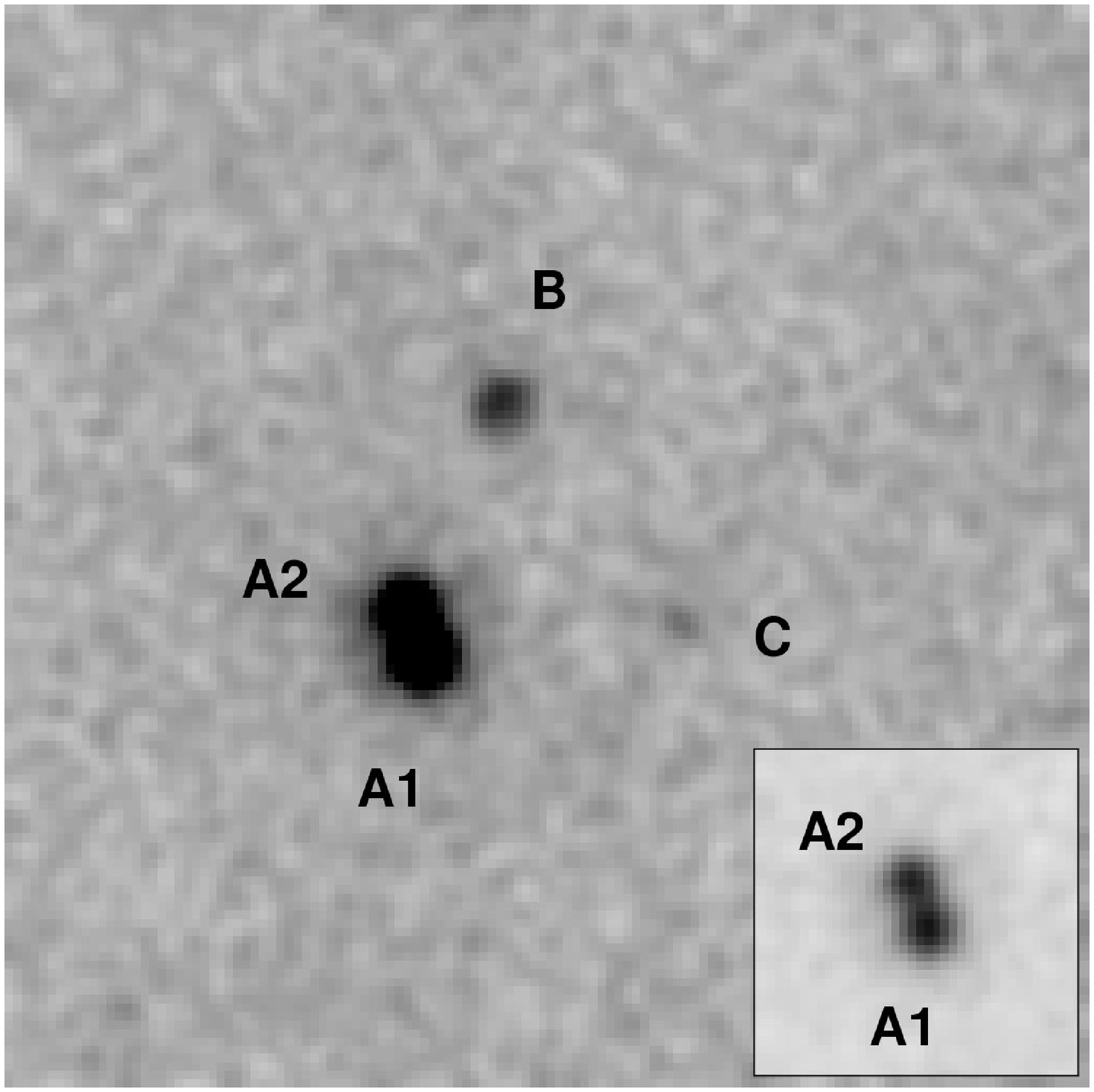}{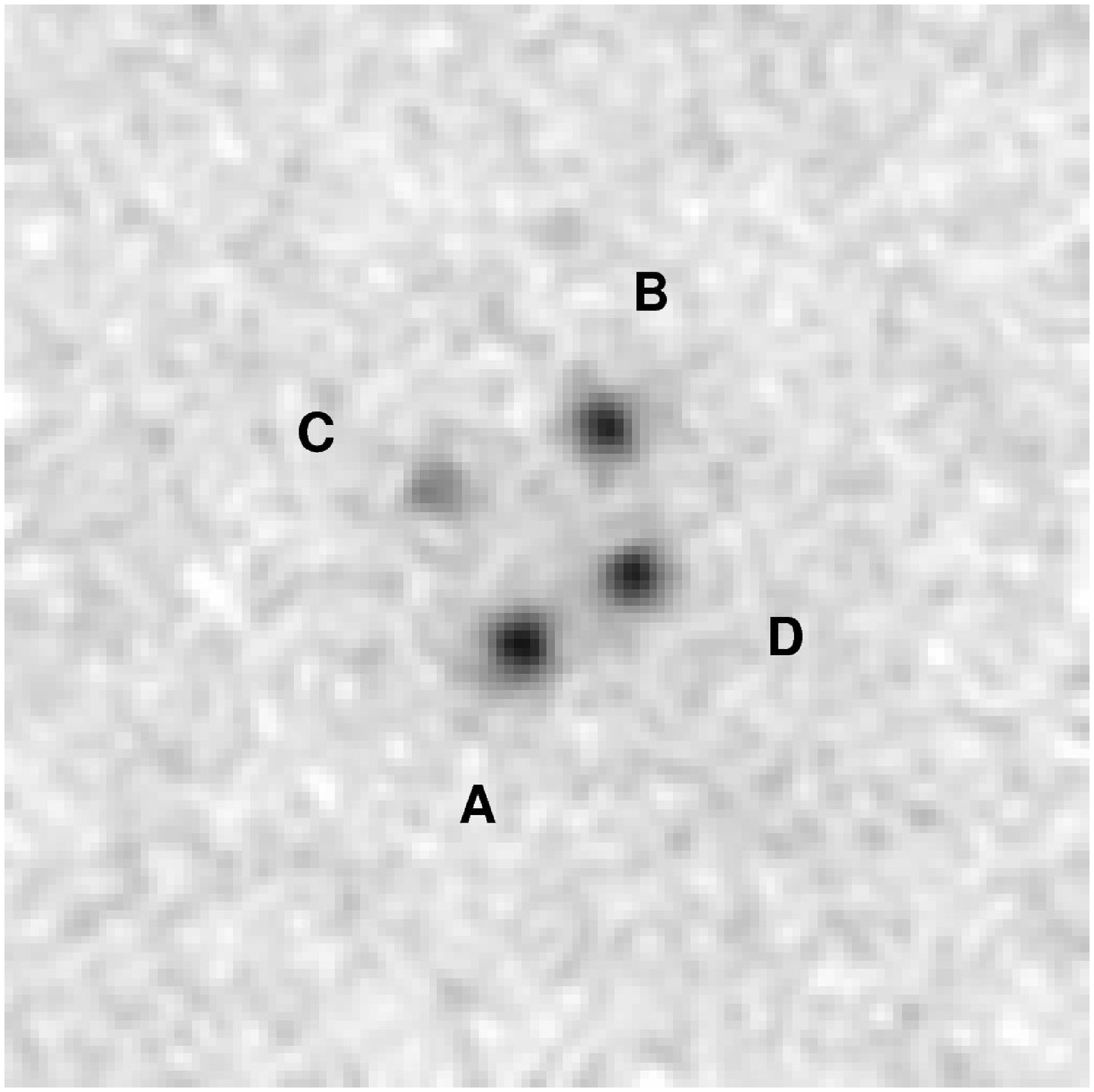}
\caption{
Quadruple lens systems MG0414$+$0534 (left) and Q2237$+$030 (right)
at $11.7\ \mu$m taken with COMICS/Subaru on 2005 October 10 and 11 (UT).
{\it Inset (left):} Image A1 and A2 of MG0414$+$0534 in
a different level of grayscale.
North is up, east is left,
and the pixel scale of them is $0.\arcsec 065$ pixel${}^{-1}$.
These images have been smoothed with
a Gaussian kernel of $\sigma = 0.\arcsec 065$
in order to improve their visual impression.
}
\end{figure*}

\section{OBSERVATION}

The mid-infrared imaging of MG0414$+$0534 and Q2237$+$030
was carried out on 2005 October 10 and 11 (UT),
using the cooled mid-infrared camera and spectrometer
(COMICS; Kataza et al. 2000) attached on
the Cassegrain focus of the Subaru telescope.
The field of view is $42\arcsec \times 32\arcsec $ and
the pixel scale is $0.\arcsec 129$ pixel${}^{-1}$.
We used the N11.7 filter, whose effective wavelength and bandwidth
are $\lambda_c = 11.67\ \mu$m and $\Delta \lambda=1.05\ \mu$m, respectively.
The chopping frequency was set $0.45$ Hz with a width of $10\arcsec$,
and in addition to the chop,
the telescope position was dithered during the observation.
The FWHM of point-spread function (PSF) was about $0.\arcsec 33$,
and the diffraction core was seen in the PSF.
The total exposure times of the available data were
3.2 hours for MG0414$+$0534 and 5.8 hours for Q2237$+$030, respectively.
The nights were photometric during the observation,
and HD25477 and HD206445 were observed
for the photometric standard stars \citep{cohen99}.

The images were reduced
in the same way as described in detail in Paper I.
The resultant mid-infrared ($\lambda = 11.7\ \mu$m) images
of the lensed QSOs, MG0414$+$0534 and Q2237$+$030,
are presented in Figure 1.
The images presented in the figure
have been re-sampled with half size of pixels
(the pixel scale is $0.\arcsec 065$ pixel${}^{-1}$).
They also have been smoothed with
a Gaussian kernel of $\sigma _r = 0.\arcsec 065$
in order to improve their visual impression,
although photometries were carried out based on the images
without any smoothing.
As presented in Figure 1,
the lensed images A1 and A2 of MG0414$+$0534
and all lensed images of Q2237$+$030 were
clearly detected and well separated from each other.
The faint lensed images B and C of MG0414$+$0534 were also detected.

\section{RESULTS}

The fluxes and the flux ratios of the lensed images
of MG0414$+$0534 and Q2237$+$030
were derived as described in Paper I.

The fluxes of $f$(A1$+$A2) of MG0414$+$0534
and $f$(A$+$B$+$D) of Q2237$+$030
were measured by aperture photometry,
then the total fluxes of all lensed images
were estimated from the aperture-photometry fluxes
and the flux ratios obtained as described later,
because the images B, C of MG0414$+$0534
and the image C of Q2237$+$030 were much fainter
than the others.
Since the lensed images were so close each other,
the fluxes of $f$(A1$+$A2) of MG0414$+$0534
and $f$(A$+$B$+$D) of Q2237$+$030
were measured by connecting the circular apertures of
$1.\arcsec 3$ in diameter for each lensed image.
The amounts of aperture corrections were estimated
by calculating the flux losses of the photometric apertures
for MG0414$+$0534 and Q2237$+$030 respectively,
based on the growth curve of flux within circular apertures
(from $0.\arcsec 5$ to $4.\arcsec 6$ in diameter)
derived from the multiaperture photometry
of the photometric standard stars.
They were estimated as about 10\% and corrected.
The fluxes of the lensed QSOs were calibrated
by comparing the fluxes of the photometric standard stars,
HD25477 and HD206445, whose fluxes were calculated
by integrating its spectral energy distribution (SED)
presented by \citet{cohen99}
with the top-hat function of $\lambda_c=11.67\ \mu$m
and $\Delta \lambda=1.05 \ \mu$m.

The flux ratios between the lensed images
were estimated by PSF-fitting photometry,
with the fitting aperture of $1.\arcsec 3$ in diameter.
The PSF was modeled by a circular Gaussian radial profile
for Q2237$+$030.
On the other hand,
the PSF was modeled by
a circular double Gaussian profile for MG0414$+$0534,
in order to fit the diffraction-core plus extended-tail
profile seen in its lensed images.
The relative positions between the images were taken from 
the CASTLES web site\footnote{http://cfa-www.harvard.edu/glensdata/}
(based on the {\it Hubble Space Telescope} (HST) imaging observations).
The free parameters of the fit were
the fluxes of the lensed images,
the positional shift of the whole images,
and the widths of the model PSFs.
We first fitted only the bright lensed images
(A1, A2 of MG0414$+$0534, A, B, D of Q2237$+$030),
then fitted the remaining images
with only the parameters of their fluxes being free.

The errors of the fluxes and the flux ratios
were estimated by the following simulation,
because they were determined
not only by photon statistics and detector noise,
but also by the fluctuation of very high sky-background
in mid-infrared imaging observation,
which remains even after the image reduction process.
Here we note that
the final image of MG0414$+$0534
was obtained by summing 114 reduced images
with an exposure time of 100 s,
and that of Q2237$+$030
was obtained by summing 210 reduced images.
First, eight independent blank-sky areas of
$8.3\arcsec \times 8.3\arcsec $ were selected
surrounding the target
in each of the reduced image
with an exposure time of 100 s.
Secondly, the artificial image of the target
was added to those blank-sky areas.
It was modeled according to the results
of the aperture and the PSF-fitting photometries:
the flux of each lensed image of the target was determined
by the aperture-photometry flux and the best-fit flux ratios
obtained by the PSF-fitting photometry,
and the radial profile of each lensed image
was a circular double Gaussian profile for MG0414$+$0534,
or a circular Gaussian profile for Q2237$+$030,
with the best-fit widths of the model PSF.
Thirdly, those ``artificial target plus blank-sky'' images
were summed in the same way as did for the real data
(summing 114 images for MG0414$+$0534
 and 210 images for Q2237$+$030)
to obtain eight artificial final images.
Finally, the aperture and the PSF-fitting photometries
were applied to them in the same way as did for the real data
to measure the fluxes and the flux ratios.
Then, the standard deviations of them were calculated
to estimate the errors of the flux and the flux ratios.

The fluxes, the flux ratios, and their 1 $\sigma $ errors
of MG0414$+$0534 and Q2237$+$030
are listed in Table 1 and 2.
For comparison, 
the flux ratios in various wavelengths reported in the literature
and the prediction of the smooth lens models
(will be described in the next section)
are listed in Table 3 and 4.
We obtained the mid-infrared flux ratios
of MG0414$+$0534 for the first time,
and obtained the most precise measurements of
those of Q2237$+$030.
These properties in the flux ratios of both lens systems
provide important implications for the nature of lens
substructure, as discussed in the next section.

\begin{deluxetable}{lllll}
\label{tab1}
\tablecolumns{5}
\tablewidth{0pt}
\tablecaption{Mid-infrared Flux and Flux Ratio of MG0414$+$0534}
\tablehead{
\multicolumn{2}{c}{flux} & \multicolumn{3}{c}{flux ratio} \\
\colhead{A1$+$A2} & \colhead{All} & \colhead{A2$/$A1} & \colhead{B$/$A1} & \colhead{C$/$A1} \\
\colhead{(mJy)} & \colhead{(mJy)} & \colhead{} & \colhead{} & \colhead{}
}
\startdata
$31.2\pm 1.0$ & $39.2\pm 1.4$ & $0.90\pm 0.04$ & $0.36\pm 0.02$ & $0.12\pm 0.03$ 
\enddata
%\tablenotetext{a}{}
\end{deluxetable}

\begin{deluxetable}{lllll}
\label{tab2}
\tablecolumns{5}
\tablewidth{0pt}
\tablecaption{Mid-infrared Flux and Flux Ratio of Q2237$+$030}
\tablehead{
\multicolumn{2}{c}{flux} & \multicolumn{3}{c}{flux ratio} \\
\colhead{A$+$B$+$D} & \colhead{All} & \colhead{B$/$A} & \colhead{C$/$A} & \colhead{D$/$A} \\
\colhead{(mJy)} & \colhead{(mJy)} & \colhead{} & \colhead{} & \colhead{}
}
\startdata
$19.0\pm 0.9$ & $22.2\pm 1.0$ & $0.84\pm 0.05$ & $0.46\pm 0.02$ & $0.87\pm 0.05$ 
\enddata
%\tablenotetext{a}{}
\end{deluxetable}

\begin{deluxetable*}{llllll}
\label{tab3}
\tablecolumns{6}
%\rotate
\tablewidth{0pt}
\tablecaption{Flux Ratios of MG0414$+$0534}
\tablehead{
\colhead{wavelength} & \colhead{date} 
                     & \multicolumn{3}{c}{flux ratio} & reference\\
                     &
                     & \colhead{A2$/$A1} & \colhead{B$/$A1} & \colhead{C$/$A1} 
}
\startdata
F814W ($0.8\ \mu $m)     &                   &$0.425\pm 0.033$&$0.474\pm 0.040$&$0.215\pm 0.018$& 1\\
F110W ($1.1\ \mu $m)     &                   &$0.637\pm 0.024$&$0.417\pm 0.016$&$0.194\pm 0.007$& 1\\
F160W ($1.6\ \mu $m)     &                   &$0.738\pm 0.025$&$0.391\pm 0.013$&$0.179\pm 0.005$& 1\\
F205W ($2.1\ \mu $m)     &                   &$0.832\pm 0.017$&$0.384\pm 0.015$&$0.174\pm 0.002$& 1\\
22 GHz                   & 1994.04.01        &$0.881\pm 0.013$&$0.397\pm 0.004$&$0.157\pm 0.003$& 2\\
15 GHz                   & 1993.01.15        &$0.8905\pm 0.0049$&$0.3896\pm 0.0015$&$0.1515\pm 0.0013$& 2\\
 8 GHz                   & 1993.01.15        &$0.8974\pm 0.0016$&$0.3887\pm 0.0005$&$0.1492\pm 0.0004$& 2\\
 5 GHz                   & 1993.01.15        &$0.898\pm 0.010$&$0.386\pm 0.003$&$0.144\pm 0.002$& 2\\
\\
11.7 $\mu $m             & 2005.10.11, 12    & $0.90\pm 0.04$ & $0.36\pm 0.02$ & $0.12\pm 0.03$ & 3\\
SIE+ES                   &                   &$1.090$         &$0.257$         &$0.107$         & 4 \\
SIE+ES+X                 &                   &$1.063$         &$0.321$         &$0.158$         & 4
\enddata
\tablecomments{The references are 1: CASTLES, 2: \citet{katz97},
3: this work, the mid-infrared observation,
and 4: this work, the lens model with smooth potential.}
%\tablenotetext{a}{}
\end{deluxetable*}

\begin{deluxetable*}{llllll}
\label{tab4}
\tablecolumns{6}
%\rotate
\tablewidth{0pt}
\tablecaption{Flux Ratios of Q2237$+$030}
\tablehead{
\colhead{wavelength} & \colhead{date} 
                     & \multicolumn{3}{c}{flux ratio} & reference\\
                     &
                     & \colhead{B$/$A} & \colhead{C$/$A} & \colhead{D$/$A} 
}
\startdata
C$_{\rm III}]$ continuum & 2002.07.16, 17    &$0.200\pm 0.036$&$0.484\pm 0.150$&$0.325\pm 0.091$& 1\\
Mg$_{\rm II}$ continuum  & 2002.07.16, 17    &$0.232\pm 0.042$&$0.406\pm 0.089$&$0.351\pm 0.070$& 1\\
C$_{\rm III}]$ line      & 2002.07.16, 17    &$0.492\pm 0.123$&$0.602\pm 0.187$&$0.624\pm 0.175$& 1\\
Mg$_{\rm II}$ line       & 2002.07.16, 17    &$0.505\pm 0.086$&$0.487\pm 0.107$&$0.566\pm 0.113$& 1\\
H$_{\beta }$ line        & 2002.08.10        &$0.376\pm 0.007$&$0.387\pm 0.007$&$0.461\pm 0.004$& 2\\
$[$O$_{\rm III}]$ line   & 2002.08.10        &$0.81\pm 0.04$&$0.88\pm 0.05$&$1.09\pm 0.05$& 2\\
8.9 \& 11.7 $\mu $m      & 1999.07.28, 09.24 & $1.11\pm 0.11$ & $0.59\pm 0.09$ & $1.00\pm 0.10$ & 3\\ 
11.7 $\mu $m             & 2000.07.11        & $1.11\pm 0.09$ & $0.72\pm 0.07$ & $1.17\pm 0.09$ & 4\\ 
8 GHz                    & 1995.06.25        & $1.08\pm 0.27$ & $0.55\pm 0.21$ & $0.77\pm 0.23$ & 5\\
\\
11.7 $\mu $m             & 2005.10.11, 12    & $0.84\pm 0.05$ & $0.46\pm 0.02$ & $0.87\pm 0.05$ & 6\\
SIE                      &                   &$0.867$         &$0.457$         &$0.855$         & 7\\
SIE+ES                   &                   &$0.887$         &$0.448$         &$0.826$         & 7
\enddata
\tablecomments{The references are 1: Wayth, O'Dowd \& Webster (2005; extinction corrected),
2: Metcalf et al. (2004; with the small aperture in their table 1),
3: \citet{agol00}, 4: \citet{agol01}, 5: \citet{falco96},
6: this work, the mid-infrared observation,
and 7: this work, the lens model with smooth potential.}
%\tablenotetext{a}{}
\end{deluxetable*}

\begin{figure*}
\figurenum{2}
\plottwo{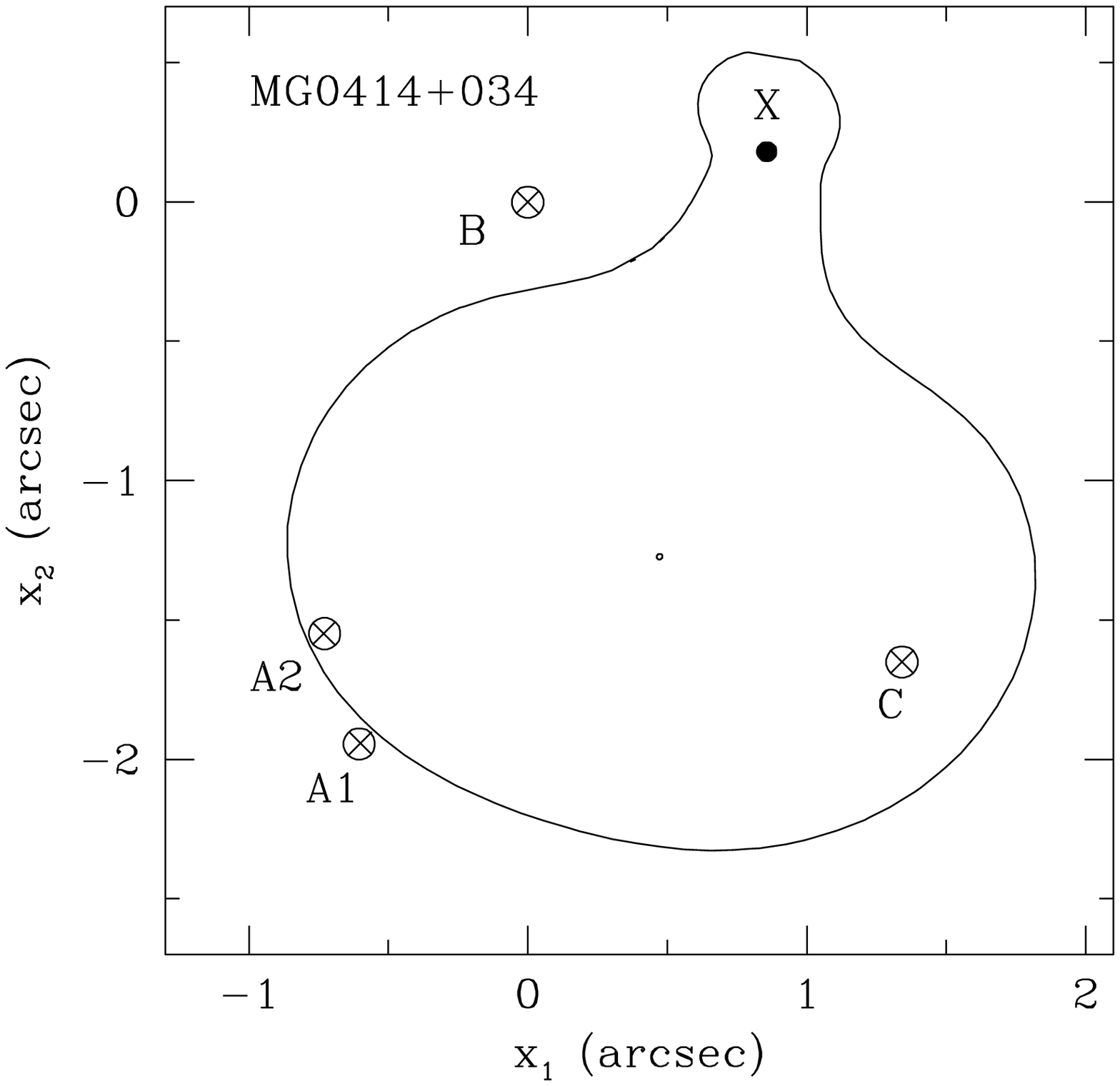}{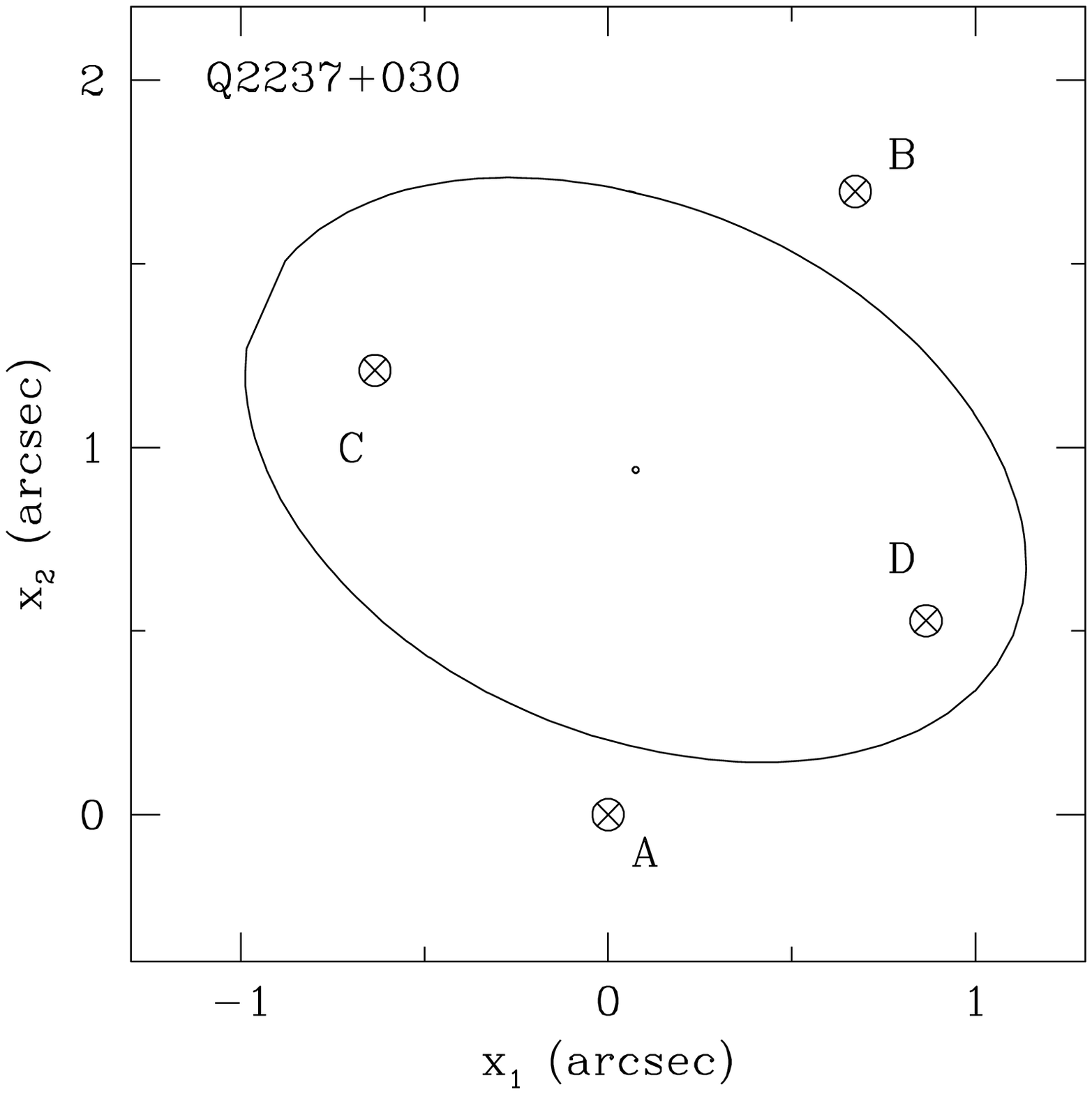}
\caption{
The best-fit model (SIE+ES+X) for MG0414$+$0534 (left) and that
(SIE+ES) for Q2237$+$030 (right). For MG0414$+$0534, the lensing effect of
a possible faint satellite (object X) is taken into account
as explained in the text. Crosses show the observed positions of
images, whereas open circles show the best-fit positions of the images
in the lens models. Solid lines denote the critical curves.}
\end{figure*}

\section{DISCUSSION}
\subsection{Flux Ratios Predicted by a Lens Model}

The configuration of these quadruple lens systems can be reproduced by
an elliptical lens. We here select a singular isothermal ellipsoid (SIE)
plus an external shear if required (SIE+ES) (Kormann, Schneider, \& Bartelmann 1994)
as a fiducial smooth lens model.
The model is a simple but sufficient representation for highlighting
the general properties of many smooth lens models which have been
constructed for reproducing these lens systems.
The current model holds the following parameters: the critical angular scale
(Einstein radius in an angular scale) characterizing the strength of
the lens potential $b$, the ellipticity $e$ of the lens and its
position angle $\theta_e$, the strength and direction of the external shear
$(\gamma,\theta_\gamma)$, and the lens galaxy position
on the lens plane $(x_0,y_0)$. The basic observational constraints are
the (relative) positions of the lensed images and lensing galaxy.
Regarding the observed mid-infrared flux ratios, we use them only for
the composite lens models of MG0414$+$0534 (i.e., considering an additional
faint satellite and/or multipoles as introduced later),
so that the models are not under-constrained by observations.
We adopt the positional data from the CASTLES web site for both systems.
For MG0414$+$0534 (Q2237$+$030), the positions relative to image B (image A)
are measured with an error of $0.\arcsec003$ ($0.\arcsec003$) for lensed images
and $0.\arcsec003$ ($0.\arcsec004$) for a lens galaxy.
We use the LENSMODEL package developed by \citet{keeton01} to find
the solutions for the image positions and undertake the $\chi^2$ fitting
to find the best model parameters, which are generally
in good agreement with previous models.
Table 5 summarizes the best model parameters and flux ratios predicted by
the current smooth lens model. 
Figure 2 shows the configuration of the best-fit lens models
for MG0414$+$0534 (left, including the effect of a faint satellite
as explained below) and Q2237$+$030 (right).
Detailed model results for each lensing system are explained in the following.

\begin{deluxetable*}{lllllllllllll}
\label{tab5}
\tablecolumns{13}
%\rotate
\tablewidth{0pt}
\tablecaption{Best Model Parameters And Predicted Flux Ratios}
\tablehead{
\colhead{model}
 & \colhead{$b$} & \colhead{$(x_0,y_0)$}
 & \colhead{$e$} & \colhead{$\theta_e$} 
 & \colhead{$\gamma$} & \colhead{$\theta_\gamma$}
 & \colhead{$N_{\rm dof}$} & \colhead{$\chi^2$}
 & \multicolumn{3}{c}{flux ratio} & \colhead{$\mu$\tablenotemark{a}} \\
\colhead{}
 & \colhead{($\arcsec$)} & \colhead{($\arcsec$)}
 & \colhead{} & \colhead{(deg)}
 & \colhead{} & \colhead{(deg)}
 & \colhead{} & \colhead{} & \colhead{} & \colhead{} & \colhead{}
}
\startdata
MG0414$+$0534 & & & & & & & & & A2$/$A1 & B$/$A1 & C$/$A1 & \\
SIE+ES\tablenotemark{b}
 & $1.174$ & $(0.449,-1.316)$
 & $0.201$ & $34$
 & $0.121$ & $-83$
 & $1$     & $297.0$
 & $1.090$ & $0.257$ & $0.107$ & $60.7$ \\
SIE+ES+X\tablenotemark{c}
 & $1.089$ & $(0.472,-1.273)$
 & $0.199$ & $-72$
 & $0.123$ & $57$
 & $0$     & $8.6$
 & $1.063$ & $0.321$ & $0.158$ & $47.0$ \\
SIE+ES+Xf\tablenotemark{d}
 & $1.088$ & $(0.473,-1.273)$
 & $0.200$ & $-72$
 & $0.124$ & $57$
 & $3$     & $24.5$
 & $1.063$ & $0.323$ & $0.158$ & $46.7$ \\
SIE+ES+Xfp\tablenotemark{e}
 & $1.065$ & $(0.470,-1.278)$
 & $0.191$ & $-72$
 & $0.145$ & $57$
 & $1$     & $10.2$
 & $1.070$ & $0.387$ & $0.133$ & $59.3$ \\
\hline
Q2237$+$030   & & & & & & & & & B$/$A & C$/$A & D$/$A  & \\
SIE
 & $0.861$ & $(0.077,0.939)$
 & $0.332$ & $67$
 &         &
 & $3$   & $17.9$
 & $0.867$ & $0.457$ & $0.855$ & $16.0$ \\
SIE+ES\tablenotemark{b}
 & $0.854$ & $(0.075,0.939)$
 & $0.371$ & $65$
 & $0.015$ & $-47$
 & $1$   & $1.0$
 & $0.887$ & $0.448$ & $0.826$ & $15.0$
\enddata
\tablenotetext{a}{Total magnification factor of all four images.}
\tablenotetext{b}{SIE plus an external shear model.}
\tablenotetext{c}{Object X modeled as an SIS with an optimized
Einstein angle of $b_X = 0.\arcsec 187$ is considered at the position of
$(0.\arcsec 857, 0.\arcsec 180)$ with an error of $0.\arcsec 010$.}
\tablenotetext{d}{The observed mid-infrared flux ratios are also used as
constraints.}
\tablenotetext{e}{Additional lens potentials with multipole terms
having an angle being aligned with $\theta_e$ are considered, given as
$|\epsilon_3|=1.298\times10^{-2}$ for $m=3$ and
$|\epsilon_4|=8.588\times10^{-3}$ for $m=4$. The observed mid-infrared
flux ratios are also used as constraints.}
\end{deluxetable*}

\subsubsection{MG0414$+$0534}

For this quadruple lens, an SIE alone yields a very large $\chi^2/N_{\rm dof}$ of
$\sim 400$ with $N_{\rm dof}=3$, so the inclusion of
an external shear (SIE+ES) is important for decreasing $\chi^2$ significantly
as listed in the table. However, this model yields still a large $\chi^2$
to accept as the best model, where some degeneracies between $e$ and $\gamma$
exist. We further consider the lensing effect of a possible faint satellite,
object X (Schechter \& Moore 1993), as has been adopted
in the previous lens models (e.g., Ros et al. 2000). For this additional
component of the lens, we assume a singular isothermal sphere (SIS)
with an optimized Einstein angle $b_X$, where the lens position is
$(x_X,y_X) = (0.\arcsec 857,0.\arcsec 180)$ with an error of
$0.\arcsec 010$ as taken from the CASTLES web site. The object X is assumed to
be at the same redshift as the lensing galaxy. We include these three parameters
of object X in our lens fitting procedure: this model is referred to as SIE+ES+X,
yielding a statistically
reasonable fit to the image positions. In addition, we consider the observed
mid-infrared flux ratios as constraints (referred to as SIE+ES+Xf
in Table 5), which however does not meaningfully improve the model fits:
the obtained $\chi^2/N_{\rm dof}\sim 8.2$ is dominated by the uncertainties in reproducing
the flux ratios ($\chi^2/N_{\rm dof}\sim 5.2$).

We compare the flux ratios predicted by these lens models with 
the observed ones, especially in the current mid-infrared wavelength of
11.7 $\mu$m and other radio wavelengths,
because the flux ratios in both wavebands are extinction free and
microlensing free, and are advantageous to search for millilensing
by CDM subhalos (e.g., Kochanek \& Dalal 2004).
We have found that the mid-infrared flux ratios agree well
with the radio flux ratios obtained by Katz, Moore, \& Hewitt (1997).
However, 
it is suggested that for MG0414$+$0534, SIE+ES and even
SIE+ES+X do not well reproduce the flux ratio of a brightest image pair
A2$/$A1 in a statistically significant manner, i.e., the observed
A2$/$A1$\la 0.90$ is systematically smaller than the best prediction
of A2$/$A1$\simeq 1.063$ by SIE+ES+X.
The deviation amounts to about 20~\% level,
or about $4$ $\sigma$ level for the mid-infrared flux ratio
and even more for the radio flux ratios.
The increase of the mass of object X or the change of its redshift
may further reduce A2$/$A1
to be consistent with the observed ratio, which however
leads to larger discrepancy with other flux ratios such as B$/$A1 and C$/$A1.
It is worth noting here that given a set of best-fit model parameters,
the change for one of those within $\Delta \chi^2 \le 1$ yields the
change of flux ratios being only $O(10^{-3})$. Thus, we need to 
adopt a more complex model in order to explain the observed flux ratios.

To further test the possibility of reproducing A2$/$A1 for MG0414$+$0534,
we consider the addition of lens potentials with higher order multipoles,
$\phi_m = (\epsilon_m r / m) \cos m(\theta-\theta_m)$, to the primary
lens model including object X, where $r$ and $\theta$ denote a projected
radius and angle, respectively, and $\epsilon_m$ and $\theta_m$ denote
the amplitude and orientation angle of $\phi_m$, respectively.
We confine our attention to $m=3$ and 4 terms, where $\theta_m$ is constrained
to be aligned with the ellipsoid $(\theta_m = \theta_e)$ as was explored
by Kochanek \& Dalal (2004).
The mid-infrared flux ratios are also used as constraints, so
$N_{\rm dof}$ is $1$ in this model (referred to as SIE+ES+Xfp in Table 5).
Inclusion of such multipoles may be able to reproduce anomalous flux ratios
(Evans \& Witt 2003). However, the discrepancy with the observed A2$/$A1
still remains; the predicted ratio A2$/$A1 of 1.070 from this model
is even worse than that without multipoles, 1.063.
We thus require an additional lens substructure possibly near these images,
whereby their flux ratio is directly modified.

\subsubsection{Q2237$+$030}

For the Einstein Cross, an SIE alone reasonably reproduces the image positions
($\chi^2/N_{\rm dof}\sim6.0$) and the consideration of an external shear
(SIE+ES) yields successful fits ($\chi^2/N_{\rm dof}\sim1.0$).
These models are also able to reproduce
the observed flux ratios in the mid-infrared wavelength
(this work; Agol et al. 2000, 2001)
and those in the radio wavelength \citep{falco96}:
the differences between the modeled flux ratios
and our mid-infrared flux ratios
are confined well within 1~$\sigma$ level.
Even if our mid-infrared flux ratios are also used
as additional constraints, we obtain the resultant
change of $\chi^2$ as only $\Delta\chi^2 \sim 0.4$ for an SIE+ES model,
which indicates that the model
fitting to the flux ratios is satisfactory. Indeed, many smooth models
for this lens system have already been constructed so far, as summarized in
Wyithe, Agol, \& Fluke (2002), and some of previous models have also reproduced
the current mid-infrared flux ratios in a reasonable manner. This suggests
that we do not necessarily require an explicit lens substructure
for the reproduction of the mid-infrared flux ratios.

\subsection{Limits on Substructure Lensing}
Within the limitation of our smooth lens models, it is yet unsuccessful
to explain the mid-infrared flux ratio of A2$/$A1$=0.90 \pm 0.04$ for MG0414$+$0534.
This appears to be also the case in other, more complicated lens models
based on a multipole-Taylor expansion for a lens potential
by Trotter, Winn, \& Hewitt (2000). They made use of high-resolution,
VLBI radio continuum maps, which show the complex spatial structure
with multiple components and kinks (Ros et al. 2000). Such detailed features
of lensed images are also a possible probe for CDM substructures
(Inoue \& Chiba 2005a,b; Chen et al. 2007). However Trotter et al.'s model
being best fitted to the positions of multiple components of the lensed images
does not reproduce the radio flux ratios well. In particular, the suggested
flux ratio A2$/$A1 is systematically larger than unity, contrary to
the observed flux ratios (see also Kochanek \& Dalal 2004).

In this subsection, we explore the presence of another lens substructure
(i.e., other than object X) in the form of a star or a CDM subhalo, which
can cause, if being located in the vicinity of the lensed images,
the discrepancy between the observed and predicted flux ratios.
The mid-infrared flux ratio is particularly useful in this regard,
because the size of the corresponding near-infrared emitting region
in the rest frame, which corresponds to
the innermost dust torus of the lensed QSOs,
can be estimated from the luminosity of the central engine.
Based on the results of dust reverberation observations
\citep{minezaki04,suganuma06}, where the lag time between the $K$-band
and UV-optical continuum flux variations is calibrated,
the radius of the near-infrared ($K$-band)
emitting region can be estimated as
$\log R\ ({\rm light\ days}) = (-2.15 - M_V / 5.0)\times c$,
where $M_V$ is the absolute magnitude of AGN in $V$ band
and $c$ is the light velocity,
or $R\ ({\rm pc}) = 1.2\times (\lambda L_{\lambda }^{\rm opt}
/10^{46}\ {\rm ergs\ s^{-1}})^{1/2}$,
where $\lambda L_{\lambda }^{\rm opt}$ is
the optical ($\lambda=0.51\ \mu m$) luminosity
\footnote{
In Paper I, the estimation of $M_V$ and resultant size of a torus
contained small errors: correctly, the radii of the rest-frame
$K$-band emission region are estimated as $R=0.4$ pc for PG1115$+$080
and $R=2.2$ pc for B1422$+$231.
Other parameters which depend on $R$ change only very slightly,
so the conclusion in Paper I remains unchanged.
} 
.

Unfortunately, it is difficult to obtain the optical luminosities
of MG0414$+$0534 and Q2237$+$030 from the fluxes
in the rest-frame optical wavelengths.
The rest-frame optical flux of MG0414$+$0534 is affected by large extinction, 
whose nature is not fully understood
(e.g., Angonin-Willaime et al. 1999; Eliasdottir et al. 2006),
and that of Q2237$+$030 is affected by microlensing
in addition to extinction
(Udalski et al. 2006 and references therein; Eliasdottir et al. 2006).
Instead, we estimate their optical luminosities
from the mid-infrared fluxes and the composite SED of QSOs.
Although the scatter of the rest-frame optical to near-infrared
luminosity ratio is not small, its systematic effect
such as the luminosity dependency and the redshift dependency
seems to be small \citep{jiang06}.

From the mid-infrared fluxes
and the composite SED of QSOs presented by \citet{richards06},
we obtain the optical luminosity of
$\lambda L_{\lambda }^{\rm opt}=9.8\ (\mu /50)^{-1}\times 10^{45}\ {\rm ergs\ s^{-1}}$
and
$\lambda L_{\lambda }^{\rm opt}=6.0\ (\mu /15)^{-1}\times 10^{45}\ {\rm ergs\ s^{-1}}$
for MG0414$+$0534 and Q2237$+$030 respectively,
where $\mu $ is the magnification factor of the lens model.
Based on the dust reverberation,
the source radii of the rest-frame $K$-band emitting region
are estimated as
$R = 1.2\ (\mu /50)^{-1/2}$ pc and $R = 0.9\ (\mu /15)^{-1/2}$ pc.
In a longer rest-frame wavelength than $K$ band,
the radius of a near-infrared emitting region is larger than that at $K$ band,
because the cooler dust distributed at larger radii of the dust torus
would contribute more to the near-infrared flux.
The rest-frame wavelengths of observing $11.7\ \mu$m emission
are $3.2\ \mu$m for MG0414$+$0534 and $4.3\ \mu$m for Q2237$+$030,
then we estimate the source radius as
$R_S\sim 2\ (\mu /50)^{-1/2}$ pc and $R_S \sim 2\ (\mu /15)^{-1/2}$ pc
for MG0414$+$0534 and Q2237$+$030, respectively.
Their angular sizes are estimated as
$\theta_S \sim 3 \times 10^{-4}(\mu /50)^{-1/2}$ arcsec
and $\theta_S \sim 2\times 10^{-4}(\mu /15)^{-1/2}$ arcsec.

To be compared with $\theta_S$ is an Einstein angle $\theta_E$ provided
by either a star with sub-solar mass (microlensing) or a CDM subhalo with
$M = 10^{7-9} M_\odot$ (millilensing), which causes anomaly in the flux ratios
of lensed images (Wyithe et al. 2002; Paper I).
Denoting $M_E$ as a substructure mass inside $\theta_E$,
the latter is expressed as
$\theta_E = 5 \times 10^{-7} (M_E/0.1 M_\odot)^{1/2}$ arcsec
for MG0414$+$0534 and
$\theta_E = 2 \times 10^{-6} (M_E/0.1 M_\odot)^{1/2}$ arcsec
for Q2237$+$030.
As a reference for a spatially extended substructure
being applicable to a CDM subhalo, we use an SIS.
For its mass indicator, we use $M_{100}^{\rm SIS}$
that is defined as the the mass within a radius of 100~pc
from the center of an SIS.
As shown in Paper I,
an SIS having a circular velocity of $V_c$ provides an Einstein radius
of $\theta_E \propto V_c^2$, while $V_c^2 \propto M_{100}^{\rm SIS}$
for a given radius of 100~pc.
We then obtain
$\theta_E = 3 \times 10^{-4} (M_{100}^{\rm SIS}/10^6 M_\odot)$ arcsec
for MG0414$+$0534 and
$\theta_E = 6 \times 10^{-4} (M_{100}^{\rm SIS}/10^6 M_\odot)$ arcsec
for Q2237$+$030.

Comparison between $\theta_E$ and $\theta_S$ indicates that a star with
sub-solar mass is unable to magnify mid-infrared flux arisen from
a dust torus, because $\theta_E \ll \theta_S$ (or $\theta_E < \theta_S/10$,
see Wyithe et al. 2002 for this argument).
For Q2237$+$030, the agreement between the observed mid-infrared
flux ratios and the predicted ones by a smooth lens model suggests
$\theta_E < \theta_S/10$, which yields $M_E < 8 (\mu/15)^{-1} M_\odot$ or
$M_E < 8 M_\odot$ using an SIE+ES smooth model with $\mu = 15.0$ (Table 5).
Thus, the reported anomalous flux ratios in the optical bands are
caused by microlensing by a star (Schneider et al. 1988; Irwin et al. 1989),
not by millilensing of a CDM subhalo.
For MG0414$+$0534, both of the radio and mid-infrared flux ratios appear to be
modified by substructure lensing compared with the prediction of
a smooth lens model; the difference in the most problematic flux ratio, A2$/$A1,
is at most about 20~\%. Taking into account $\theta_S$ of a dust torus,
the change of a flux ratio by at most about 20~\% from a smooth-lens prediction
can be caused by a substructure with $\theta_E \simeq \theta_S/10$
if its position is just centered at either of the lensed image A2 or A1
\citep{inoue05b}; a more massive substructure located outside the image
can give rise to the same amount of a flux perturbation.
Thus, we obtain a lower limit on $M_E$, namely,
$M_E \ga 340 (\mu/50)^{-1} M_\odot$ inside an Einstein angle
(i.e., $M_E \ga 360 M_\odot$ for an SIE+ES+X model with $\mu = 47.0$),
thereby suggesting a CDM subhalo is likely to affect the mid-infrared
flux ratio. If it is modeled as an SIS, we obtain
$M_{100}^{\rm SIS} \ga 9.7 \times 10^4 (\mu/50)^{-1/2} M_\odot$
inside radius of 100~pc
(i.e., $M_{100}^{\rm SIS} \ga 1.0 \times 10^5 M_\odot$
for an SIE+ES+X smooth model).
\\

\begin{figure}
\figurenum{3}
\plotone{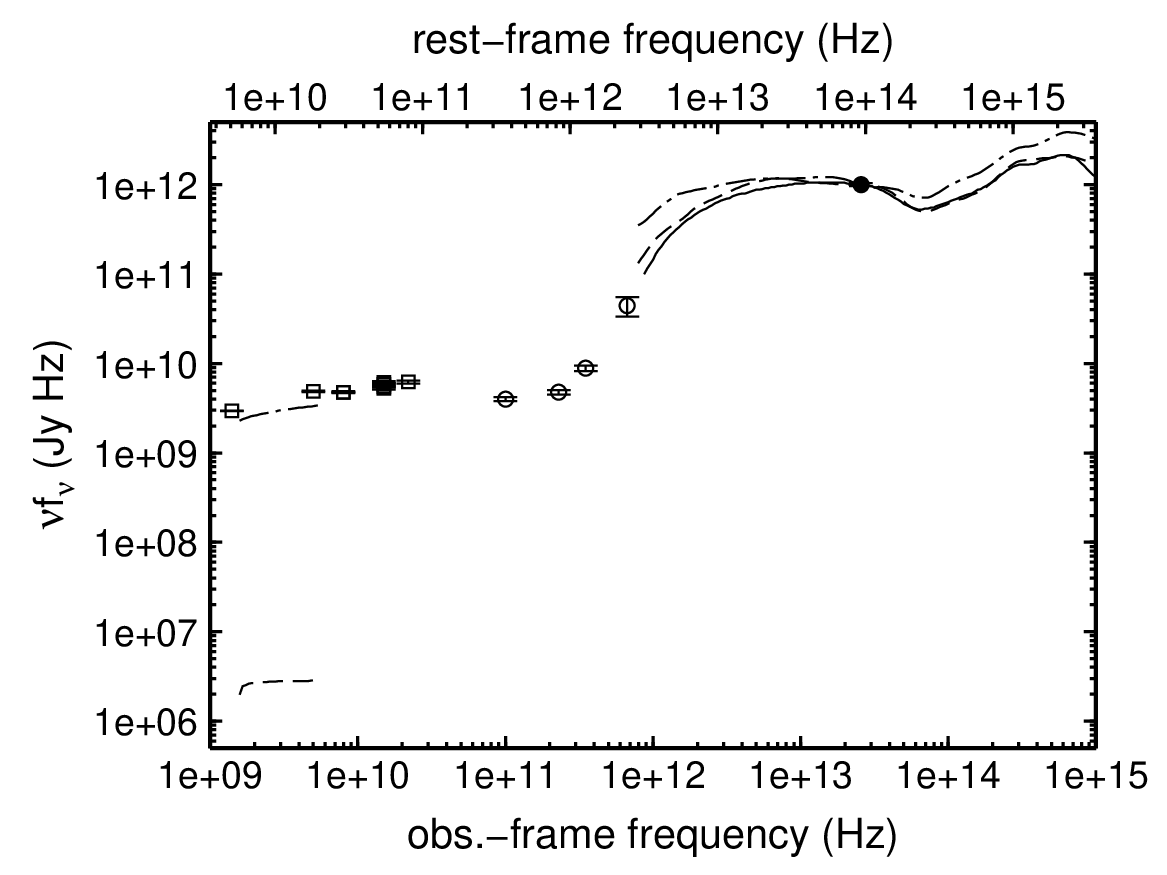}
\caption{
The spectral energy distribution (SED) of MG0414$+$0534
from infrared to radio wavelength.
The filled circle is the $11.7$ $\mu $m flux obtained by this work.
The open circles are the sub-mm fluxes taken from \citet{barvainis02},
and the open boxes are the radio continuum fluxes taken from \citet{katz97}.
The solid line is the composite SED of QSOs by \citet{richards06},
the dashed and dash-dotted lines are, respectively,
the mean energy distributions (MEDs) of
radio-quiet and radio-loud QSOs by \citet{elvis94}.
}
\end{figure}

\begin{figure}
\figurenum{4}
\plotone{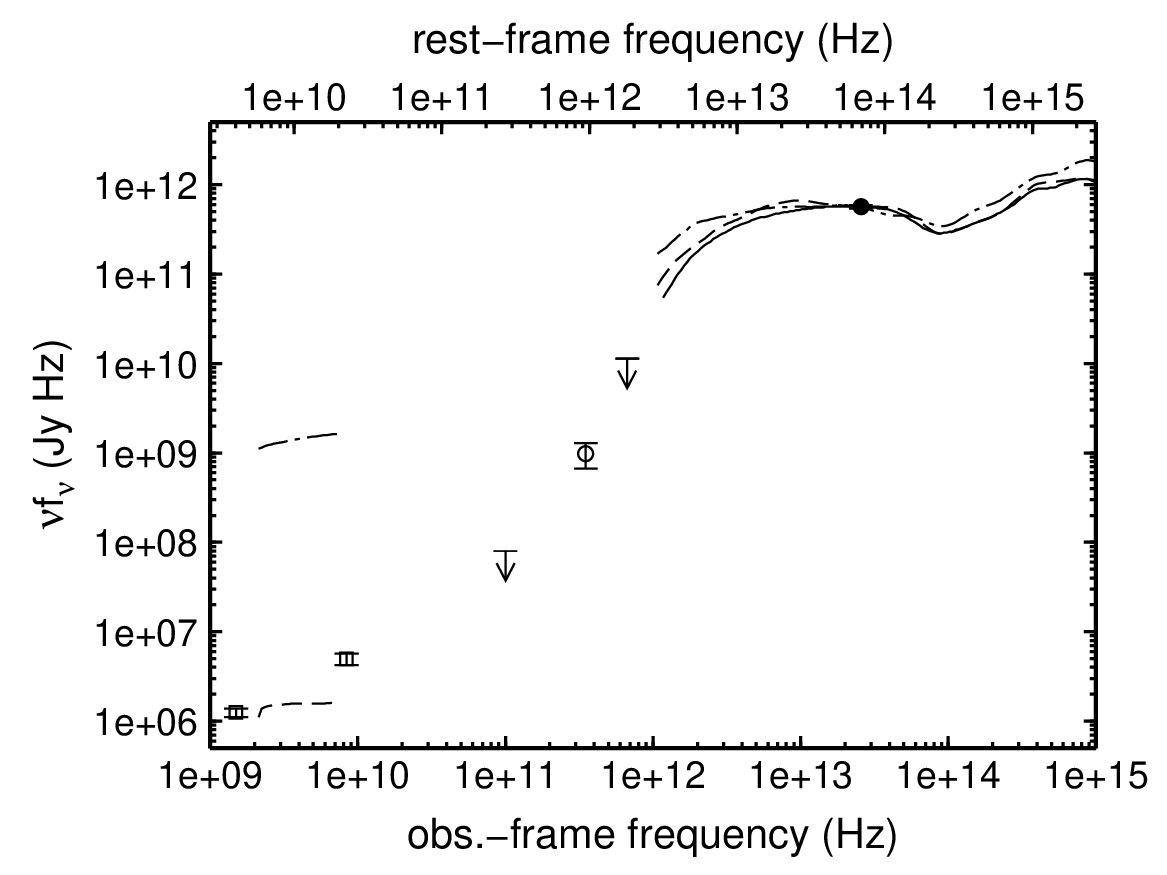}
\caption{
The SED of Q2237$+$030
from infrared to radio wavelength.
The filled circle is the $11.7$ $\mu $m flux obtained by this work.
The open circle and the upper limits are the sub-mm fluxes
taken from \citet{barvainis02}, and
the open boxes are the radio continuum fluxes
taken from \citet{falco96}.
The solid line is the composite SED of QSOs by \citet{richards06},
the dashed and dash-dotted lines are, respectively,
the MEDs of radio-quiet and radio-loud QSOs by \citet{elvis94}.
}
\end{figure}

\subsection{Other Possible Mid-Infrared Sources}
Although infrared flux of AGN is dominated by 
thermal emission of the dust torus,
a minor fraction might be contributed by other sources,
such as an extension of optical continuum emission
from the accretion disk (Kishimoto, Antonucci, \& Blaes 2005;
Tomita et al. 2006; Minezaki et al. 2006), non-thermal emission
from a compact region relating to radio activity
\citep{Neugebauer99,enya02},
or infrared emission from starburst activity, which is 
suggested for some high redshift QSOs \citep{carilli01,omont01}.
Since the former two sources are more compact than the dust torus
while the latter is more spatially extended,
possible contributions of both of them
to the mid-infrared flux
might have an influence on the estimation
of the limits on the substructure.
We will examine their possible contributions.

The source size of the accretion disk is so small
that the flux originated from it could be
affected by microlensing.
However, the mid-infrared flux ratios of
both of MG0414$+$0534 and Q2237$+$030
are considered not to be affected by
the microlensing of the accretion disk component.
Firstly,
the mid-infrared flux ratios and the radio flux ratios
of MG0414$+$0534 are in good agreement with each other.
The source size of the radio emission is
also considered to be so large that its flux
would not be affected by microlensing (e.g., Kochanek \& Dalal 2004),
and the agreement with both flux ratios
suggests that the mid-infrared flux ratios of MG0414$+$0534
are not affected by microlensing.
Secondly, possible time variation of mid-infrared flux ratios
of Q2237$+$030 at an interval of $5-6$ years are examined
by comparing \citet{agol00,agol01} with this work.
We found that the mid-infrared flux ratios are almost constant,
or the small change between them shows an opposite trend to
that observed in the optical flux ratios that are heavily 
affected and changed by microlensing (see Section 4.4 for detail).
The mid-infrared flux ratios of Q2237$+$030 
obtained in different epochs of observations
suggest that they are also not affected by microlensing.

The contribution of non-thermal emission
to the mid-infrared flux is estimated
based on the SEDs of MG0414$+$0534 and Q2237$+$030
from infrared to radio wavelengths.
Their SEDs are presented in Figure 3 and Figure 4.
The filled circles in the figures are the $11.7$ $\mu $m flux
obtained by this work.
The open circles and the upper limits are the sub-mm fluxes
taken from \citet{barvainis02},
and the open boxes are the radio continuum fluxes
taken from \citet{katz97} for MG0414$+$0534
and \citet{falco96} for Q2237$+$030, respectively.
The solid line is the composite SED of QSOs by \citet{richards06},
the dashed line and the dash-dotted lines are
the mean energy distribution (MED) of
radio-quiet QSOs and radio-loud QSOs by \citet{elvis94}.
Those composite SEDs of QSOs are normalized
at the $11.7$ $\mu $m fluxes.
As presented in Figure 3 and Figure 4 respectively,
the infrared to radio SED of MG0414$+$0534 agrees with
the typical SED of radio-loud QSOs,
while that of Q2237$+$030 agrees with
the typical SED of radio-quiet QSOs.

It is generally accepted that
the infrared emission ($\sim 10^{11.5-14.5}$ Hz) of QSOs
is dominated by thermal reradiation of dust torus
heated by UV radiation from the accretion disk,
while the radio emission ($\lesssim 10^{11.5}$ Hz) of QSOs
is non-thermal in origin, synchrotron radiation
from relativistic jet.
Since the radio continuum emission of MG0414$+$0534
obtained by \citet{katz97} follows a power-law SED,
the contribution of non-thermal emission
to the mid-infrared flux of MG0414$+$0534
is estimated by simply extrapolating the power-law SED
of the radio continuum emission to the mid-infrared wavelength.
Although MG0414$+$0534 is a radio-loud QSO,
it is estimated to less than $\sim 5$\%,
and moreover, it would be smaller than this estimation,
because the power-law SED by synchrotron radiation would have
a break at some frequency higher than the radio wavebands.
The radio to mid-infrared flux ratio of Q2237$+$030
is about 3 order of magnitudes smaller than
that of MG0414$+$0534 as presented in Figure 3 and Figure 4,
thus the contribution of non-thermal emission 
to the mid-infrared flux of Q2237$+$030
is expected to be quite small.
Therefore, 
the possible non-thermal emission in the mid-infrared flux
would have little effect on the discussion of
substructure lensing for both targets.

In addition to AGN activity, starburst activity
is known to be another important source for powerful
far-infrared emission of galaxies (e.g. Sanders \& Mirabel 1996).
\citet{carilli01} and \citet{omont01} carried out
observations at 250 GHz (1.2 mm) of
a total of $\sim 100$ high redshift ($z\gtrsim 4$) QSOs.
About one third of them were detected at 250 GHz,
which showed excess emission over the composite SED
of low-redshift radio-quiet QSOs in rest-frame far-infrared wavelength.
They suggested that substantial fraction of the far-infrared luminosity
arises from starburst activity,
although they could not exclude the possibility of its AGN origin.
Thus, starburst activity associated with QSOs
could be assessed by the excess emission in far-infrared wavelength.
As presented in Figure 3 and Figure 4,
the far-infrared fluxes of MG0414$+$0534 and Q2237$+$030
fall at the MED 
(and its interpolation at far-infrared to sub-mm wavelengths)
of radio-quiet QSOs normalized by their mid-infrared fluxes,
and no excess emission in far-infrared is detected.
Therefore, they do not host significant starburst,
and the contribution of starburst activity
would be unimportant.

\subsection{Comparison with Other Observations of Q2237$+$030}
The mid-infrared flux ratios of Q2237$+$030
obtained by \citet{agol00,agol01} and those of this work
are almost the same within the errors.
However, the flux ratio of image A is different from 
Agol et al. (2001; $f({\rm A})/f({\rm total})=0.250\pm 0.015$)
and this work
($f({\rm A})/f({\rm total})=0.315\pm 0.005$)
at about $4\sigma $ level,
and sorting the images in order of brightness,
$f({\rm B}) \approx \ f({\rm D})\gtrsim f({\rm A})>f({\rm C})$
by \citet{agol01} while
$f({\rm A})\gtrsim f({\rm B}) \approx \ f({\rm D})>f({\rm C})$
by this work.
We will discuss this difference,
which would suggest a time variation of
the mid-infrared flux ratios at an interval of five years.

First, we argue that the intrinsic variation of QSO and
the time delay between the lensed images
cannot explain the difference of the mid-infrared flux ratios
between the two observations.
That is because the lens models for Q2237$+$030 yield
the time delays between the images as less than a day
(e.g., Schneider et al. 1988; Rix et al. 1992;
 Wambsganss \& Paczynski 1994)
and they are very short compared to the timescale of
flux variation of QSOs.

If the contribution to the mid-infrared flux
originated from compact regions like the accretion disk is large,
the mid-infrared flux and flux ratio
can show time variation due to microlensing,
in a similar way to the optical flux and flux ratios.
However, according to the optical light curves presented in
\citet{udalski06},
the optical flux of image B relative to the others' fluxes
at the observation date of \citet{agol01}
was fainter than that at the date of this observation.
This is an opposite trend to the changes
in the mid-infrared flux ratios.
Therefore, the microlensing of the compact source
cannot explain the difference of the mid-infrared flux ratios
between the two observations. Thus, the origin of the 
discrepancy remains unknown. 

As shown in Table 4, a notable difference between the narrow line region
(NLR) $[$O$_{\rm III}]$ and the 
mid-infrared flux ratios of Q2237$+$030, especially for C$/$A,
also needs to be understood (Metcalf et al. 2004). This
can be explained as follows. The first possibility is that 
the NLR is so large that some of the flux 
of image A or D could leak out of the aperture to 
image C, causing the high C-to-A magnification ratio for $[$O$_{\rm III}]$.
The second possibility is that the NLR has a complex shape with
a non-uniform emissivity profile so that the total magnification
differs from that of the smaller mid-infrared emission \citep{yonehara06}.  
For instance, a double corn-like structure in the NLR
could account for the observed double peaks in image A in $[$O$_{\rm III}]$. 
We note that if any substructures (as suggested by Metcalf et al. 2004)
are present in Q2237$+$030, their impact on the mid-infrared flux
remains minor because our lens model with a smooth potential (SIE+ES)
successfully reproduces the mid-infrared flux ratios as well as
the image positions.

\section{CONCLUSION}

We have presented mid-infrared imaging at 11.7 $\mu$m for the quadruple lens
systems, MG0414$+$0534 and Q2237$+$030.
The lensed images of these QSOs have
successfully been separated from each other owing to the diffraction-limited
resolution of the COMICS/Subaru images with ${\rm FWHM}=0.\arcsec 33$.
Our calibration of the mid-infrared fluxes and flux ratios for different
lensed images have revealed that
(1) the mid-infrared flux ratio A2$/$A1 of MG0414$+$0534 shows a statistically
significant deviation from the prediction of a smooth lens model,
which is comprised of an SIE plus an external shear and a possibly
faint satellite, object X, and that
(2) the mid-infrared flux ratios between all the four images
of Q2237$+$030 are in good agreement with the prediction of 
a smooth lens model.
These results, combined with the estimated dimensions of dust tori around
these QSOs based on the relation of dust reverberation mapping,
set limits on the mass of a lens substructure, which is responsible for
the anomalous flux ratios. For MG0414$+$0534, the mass of a supposed
substructure inside an Einstein radius, is $M_E \ga 360 M_\odot$,
thus indicating that a CDM subhalo is most likely affecting the mid-infrared
flux ratio. If it is an SIS, the mass inside radius of 100~pc is given as
$\ga 1.0 \times 10^5 M_\odot$.
For Q2237$+$030, assuming a smooth lens potential, 
the observed mid-infrared
magnification flux ratios suggest $M_E < 8 M_\odot$. Thus, we conclude that 
the optical anomalous flux ratios and the time variations are entirely due to 
microlensing by stars, in agreement with Agol et al. (2000).

%%%%%%%%%%%%%%%%%%%%%%%%%%%%%%%%%%%%%%%%%%%%%%%%%%%%%%%%%
\acknowledgments
We thank Takuya Fujiyoshi for his expert assistance
during our observing runs with COMICS/Subaru.
This work has been supported in part by a Grant-in-Aid for
Scientific Research (17540210, 20340039) of the Ministry of Education, Culture,
Sports, Science and Technology in Japan.

%%%%%%%%%%%%%%%%%%%%%%%%%%%%%%%%%%%%%%%%%%%%%%%%%%%%%%%%%
%%%%%%%%%%%%%%%%%%%%%%%%%%%%%%%%%%%%%%%%%%%%%%%%%%%%%%%%%

%%%%%%%%%%%%%%%%%%%%%%%%%%%%%%%%%%%%%%%%%%%%%%%%%%%%%%%%%

%%%%%%%%%%%%%%%%%%%%%%%%%%%%%%%%%%%%%%%%%%%%%%%%%%%%%%%%%

\end{document}